\documentstyle[12pt,aasms4]{article}
\def \yskip{\penalty-50\vskip3pt plus 3pt minus 2pt}
\def \pp{\par \yskip \noindent \hangindent .4in \hangafter 1}
\def \abc#1#2#3#4 {\pp#1, {\sl#2}, {\bf#3}, #4}
\def \blank {\lower 5pt\hbox to 0.75in{\hrulefill}}
\def \kms{km~$\rm{s}^{-1}$}
\def \cc{$\rm{cm}^{-3}$}

\def \mum{$\mu$m}
\def \lsol{L$_{\odot}$}

\newfont{\rten}{cmr10} 
\def\arcdeg{\hbox{$^\circ$}}
\def\arcmin{\hbox{$^\prime$}}
\def\arcsec{\hbox{$^{\prime\prime}$}}

\begin{document}

\normalsize
 
\title{Highly Collimated Molecular Hydrogen Jets Near IRAS 05487+0255: 
NIR Imaging and Spectroscopy.}
\vspace*{0.3cm}

\author{Peter M. Garnavich}
\affil{Harvard-Smithsonian Center for Astrophysics, Cambridge, MA 02138}
\affil{email: pgarnavich@cfa.harvard.edu}
\author{Alberto Noriega-Crespo}
\affil{Infrared Processing and Analysis Center, CalTech-JPL, Pasadena, 
CA 91125}
\affil{email: alberto@ipac.caltech.edu}
\affil{Alejandro C. Raga}
\affil{Instituto de Astronom{\'{\i}}a, UNAM, 04510, M\'exico D.F., M\'exico}
\affil{email: raga@astroscu.unam.mx}
\affil{and}
\affil{Karl-Heinz B\"ohm}
\affil{Department of Astronomy FM-20, University of Washington, Seattle, 
WA 98195}
\affil{email: bohm@astro.washington.edu}

\begin{abstract}

We present new narrow-band near-infrared images together with
K band spectra of highly collimated bipolar jets close to the
IRAS 05487+0255 source. The jets are located at 
$\sim 50$\arcsec~ West of the Herbig-Haro 110 outflow.
The jets are not visible at optical wavelengths, and
therefore, do not fall into the `standard' Herbig-Haro object classification 
scheme. Nevertheless, they belong to an ever growing group of molecular 
hydrogen jets associated with YSOs which are optically undetected.
The jets are very well collimated, with a length-to-width ratio 
$\sim 10-20$. 

The spectra of the jet and counter-jet in the K-band show a limited number
of H$_2$ emission lines which makes it difficult to obtain an 
accurate excitation temperature.
We estimate $T_{ex} = 1104\pm 67$ K and $T_{ex} = 920\pm 156$ K for 
the red and blue jet components respectively.
The radial velocities of the jet and counter-jet, based on the shift of the 
(1,0) S(1) 2.121~\mum\ line, are $\sim -275\pm 50$ \kms\ and 
$\sim 180\pm 50$ \kms\ respectively, suggesting an angle of 
$\sim 30\arcdeg~-~45$\arcdeg~between the jet and the line of sight. 
The H$_2$ emission of
the entire jet extends for at least 40\arcsec~or $\sim 0.1$ pc at the
distance of Orion. If the flow velocity is comparable to that of the
radial velocities, then the dynamical age of the system is quite short 
($\sim 500$ yrs), consistent with a young jet arising from an embedded 
source. Entrainment in a turbulent mixing layer may explain this morphology
and spectral character.
\end{abstract}

\begin{keywords}
{ISM:~individual (IRAS 05487+0255)---
ISM: jets and outflows---
infrared: ISM: lines and bands---
stars: formation}
\end{keywords}

\section{Introduction}

Near infrared observations ($\sim 1-4$ \mum) are discovering a growing number
of molecular hydrogen outflows from young stellar objects (YSOs) that, 
in many cases are not detected at optical wavelengths.
Some of the most spectacular examples of these objects are 
Herbig-Haro (HH) 211 \markcite{mcc94} (McCaughrean et al. 1994) and 
L1448 \markcite{dcj94c}(see e.g. Davis et al. 1994c, and references 
therein). 
For a few objects the optical and H$_2$ jets are detected and display 
a similar morphology, as e.g. in HH 111 \markcite{gred93}
(Gredel \& Reipurth 1993) and HH 1-2 \markcite{dcj94a} (Davis et al. 
1994a;\markcite{nori94a} Noriega-Crespo \& Garnavich 1994a).
The H$_2$ jets seem to have properties intermediate to those of
the collimated optical jets and the broader CO molecular outflows:
they resemble the jets, but they trace warm  ($\sim 2000$ K) molecular gas.
Because of this they could provide important clues on the interdependence
between the atomic and molecular bipolar outflows, at least in the objects
where they appear simultaneously. Furthermore, the emission from the
H$_2$ jets could be driven by the dissipation of energy in a turbulent
mixing layer \markcite{can91} (Cant\'o \& Raga 1991) and not by shocks, where
the emission arises at the recombination regions.

In this paper we report the discovery of yet another H$_2$ jet invisible
at optical wavelengths found during our recent investigation of the 
kinematics and structure of the HH 110 \markcite{nori96} (Noriega-Crespo et 
al. 1996). The presence of the jet was already inferred from earlier
narrow band H$_2$ 2.121 \mum\ images of the HH110 region 
by \markcite{dcj94b}Davis et al. (1994b), 
but given their signal-to-noise and resolution it was impossible to 
say anything concrete about its structure. They identified two
bright K-band sources some 50\arcsec\ west of HH110 knot-A and
labeled the southern and northern emission IRS1 and IRS2 respectively.
K-band imaging of IRS1/2 was also performed by \markcite{bo91}
Reipurth and Olberg (1991) while searching for the energy source of
HH110. They matched IRS1 with the far-infrared source
IRAS~05487+0255 based on its positional coincidence, but concluded
that it was very unlikely to be the source that drives the HH110 jet. 

This paper presents improved near infrared images of the
new jets associated with IRS1 and IRS2 as  well as spectroscopic
observations in the 2\mum\ band of both the IRS2 jet and counter-jet.
The high degree of collimation of the molecular hydrogen emission may be
indicating a different excitation mechanism than those we have considered
in the past (as e.g in internal working surfaces). We have estimated the jet
radial velocities and reddening from their spectra. The observations are
reported in \S 2 and the results are summarized in \S 3.

\section{Observations}
\subsection{Imaging}

The near infrared images were obtained at the 3.5m telescope at Apache Point
Observatory on January 14 - 16, 1996, with GRIM II imager which has a 
RISC 256$\times$256 detector array. The $f/5$ optics provide a scale of 
0.\arcsec{48} per pixel and a field of view of two arcminutes. 
The filters were centered at the wavelength of the molecular hydrogen 
v = 1$-$0 S(1) 2.121 \mum~line and nearby continuum at 2.22 \mum, with 
FWHM transmission of 1\% and 4\% respectively.
 
The 2.12\mum\ images were created by a mosaic of 18 spatially contiguous
frames, with integration times of 180 seconds per frame. The continuum 
image (2.22 \mum ) was created from 12 integrations of 40 seconds
each. The FWHM resolution was between 1\arcsec~$-$1.\arcsec~5. 
The data were reduced in the following way.
Mean dark frames with an exposure times corresponding to the science
images were subtracted from each exposure and the result divided by
normalized flat field frames constructed from images of the illuminated dome.
A sky frame was created by removing the bright emission objects from each 
data frame (substituting the average local sky) and taking a median of the 
images after adjusting for slight differences in the overall sky level. 
The sky frame was then scaled and divided into each data image. 
The images were then magnified, shifted by an integer number of magnified
 pixels to align the stellar images and combined using a sigma-clipping 
algorithm. The final image was then demagnified to the same scale as the 
original frames.

\subsection{Spectra}

A K band spectrum of the IRS2 jet was obtained with the Rieke FSPEC IR 
spectrometer (256$^2$ NICMOS detector) and MMT on 1996 February 10.
A 75 line/mm grating was used that 
provided spectral coverage from 1.98 to 2.42 \mum\ with a resolution of 
34\AA\ FWHM. The slit was aligned with the jet axis at a PA=$-5^\circ$
and alternate exposures were chopped between the source and the local sky.  
The atmospheric absorption and sensitivity variation along the dispersion 
was corrected by observing bright late-F type stars at nearly the same 
airmass as the target data. The wavelength solution was determined using
the OH sky lines from the raw images. Longward of 2.3\mum\ there are few
bright sky lines and the accuracy of the calibration is reduced at the red
end of the spectrum. Since spectra of both the north and south sides of 
the jet were obtained on the same exposures, the relative velocities between
the jet and counter-jet are limited by the quality of the line centroids. 
For the strong 2.12 \mum\ line this gives an uncertainty of about 50 \kms . 

The spectra were obtained with a 1.0\arcsec\ 
wide slit, and the extraction was done over an area of 4.0 and 7.2 square
arcseconds for the north and south sections of the jet respectively. 
The line intensities were computed with these areas, and these were used to 
obtain the H$_2$ column densities by applying 
$I(v^\prime J^\prime) = {hc \over 4\pi} {\bar \nu} A(v^\prime J^\prime) 
N(v^\prime J^\prime)$, where $I(v^\prime J^\prime)$ is the line intensity,
${\bar \nu}$ the wavenumber \markcite{dab84}(from Dabrowski 1984) 
and $A(v^\prime J^\prime)$  the transition probability 
\markcite{turn77}(from Turner et al 1977). Table 2 contains the 
fluxes, intensities, and column densities for those lines 
detected in both the red-shifted North (top) and blue-shifted South (bottom)
jet components.

\section{Discussion}

The H$_2$ 2.121\mum\ image around the IRAS 05487+0255 region
is shown in Figure 1.  
The astrometry was performed under the assumption that the optical 
positions for HH 110 A, B and E along with that from reference star 1
\markcite{bo91}(Reipurth \& Olberg 1991) are known.
IRS2 is seen to be a parabolic-shaped nebula surrounding a narrow jet (11)
extending to the south.  This new H$_2$ jet also has northern component (9)
which can be traced for more than 40\arcsec\ from a dark lane between the 
jets (10) that is likely obscuring the driving source. 
The continuum-subtracted image (Figure~2) shows only the jets, proving that 
we see the jets emit narrow molecular lines while the parabolic nebula is 
dominated by scattered continuum light.

The position for the IRAS 05487+0255 source and its {\it uncertainty ellipse}
are also shown in Figure~1. The uncertainty ellipse encloses the true 
position of the source with a 95\% confidence \markcite{iras88} 
(Beichman et al. 1988). 
If this is the case then IRAS 05487+0255 does not seem to be driving any 
obvious outflow in the region and is not IRS1 as surmised by 
\markcite{bo91} Reipurth and Olberg (1991). Near the published position of
IRAS 05487+0255,  $\alpha_{1950}$ = 5$^h$ 48$^m$ 46$^s.4$  
$\delta_{1950}$ = 2\arcdeg~55 \arcmin~10\arcsec, there is a faint source 
marked (7), with $\alpha_{1950}$ = 5$^h$ 48$^m$ 47$^s$.03 $\delta_{1950}$ = 
2\arcdeg~55\arcmin~11\arcsec.1. The similarity between their positions could
 be more than a curious and fortuitous coincidence.

A third jet arises from the bright stellar source IRS1 (15), where 
there is a beam-like emission within  $\sim 3$\arcsec~from the star.
Aligned with this emission are two fainter knots 15\arcsec\ and 20\arcsec\ 
farther south (12 and 13). There is no evidence of a counter-jet to the 
north of IRS1. 

The K-band spectra of the northern and southern IRS2 jets are shown in 
Figure 3. The wavelength of the expected H$_2$ lines in this band are marked,
but given its signal-to-noise only a few lines can be considered as 
unambiguously detected (see Table 2). We have used these lines to determine 
the excitation temperature of the warm molecular gas, and to measure the 
radial velocities of both jet components.

The excitation temperature is obtained by assuming, as a first 
approximation, 
that the H$_2$ rotational-vibrational levels have equilibrium populations 
following a Boltzmann distribution. The excitation energy or temperature 
$T_{ex}$ then comes from 
$N(v^\prime J^\prime)\propto exp\lbrack {E(v^\prime J^\prime)\over 
k T_{ex}}\rbrack$, 
and such that $ln (N(v^\prime J^\prime)/g) \propto T_{ex}^{-1}$, where
$g$ is the statistical weight.
The linear least square fit to the six lines for both jet
and counter-jet (Table 2 and Figure 4) yields the following values
$T_{ex} = 1104\pm 67$ and $T_{ex} = 920\pm 156$
for the blue and red-shifted jets respectively. These temperatures
are below the excitation temperatures measured in other HH objects 
(Gredel 1994) of $\sim 2000$ K.
The flux of the 2-1 S(1) 2.24~\mum~line is uncertain, but permits
an estimate of the vibrational excitation temperature, and considering
only the v = $1-0$ vibrational transitions (without the v = $2-1$ line)
would be a measurement of the {\it rotational} excitation temperature.
These results reflect, however, the difficulty in performing such
calculations using few lines and the need for higher quality
spectra.

For some time it has been a goal of many of us to try to determine the 
dominant excitation mechanism of H$_2$ in these objects, with UV 
(fluorescence) or collisional (shock) excitation, as the most likely 
competing processes \markcite{wol91} (see e.g. Wolfire \& K\"onigl 1991). 
The most direct way to  differentiate between these mechanisms has been 
to use the 2$-$1 S(1) 2.247\mum\ to 1$-$0 S(1) 2.121 \mum\ ratio. 
The models predict ratios with values of $\sim 0.1$ for collisions and 
$\sim 0.5$ for UV excitation. From the values in Table 2 for the jet
we obtain a ratio 
of $0.13 \pm 0.05$, i.e. consistent with collisional excitation.
We mentioned before, however, that it is possible that some important 
fraction of the H$_2$ emission arises from turbulent mixing layers. 
If this is the case then the traditional H$_2$ diagnostic line ratios could 
be misleading although the emission in a turbulent mixing layer 
comes from collisionally ionized gas, detailed models of the emitted 
spectrum due to this process remain to be done \markcite{tay95} (see e.g. 
Taylor \& Raga 1995).

The spectral resolution of the IR spectra is low, nevertheless it is clear 
that lines are shifted with respect to their rest wavelengths. 
The direction of the line offsets is consistent with the  north jet 
component being red-shifted, while the south jet component is blue-shifted. 
The highest signal-to-noise line is the  v = 1$-$0 S(1) transition at 
2.121 \mum\ which gives radial velocities of $180\pm 50$ \kms and 
$-270\pm 50$ \kms\ for the northern and southern components respectively. 
If the warm molecular hydrogen gas has a flow velocity similar to the range
of the velocities measured in optical jets, $\sim 200 - 400$ \kms, then
the radial velocities suggest an angle of $\leq 45$\arcdeg, between the jet
and the line of sight.

The detectable H$_2$ emission of of both jet components extends for at least 
40\arcsec~or $\sim 0.1$ pc (at the distance of Orion for a 45\arcdeg~angle).
If the flow velocity of the molecular gas is comparable to that of the
radial velocities, $\sim 200$ \kms, then the dynamical age of the system 
is quite short, $\sim 500$ yrs. This age is consistent with the idea of
a young jet arising from an embedded source, as it seems to be the case
in HH 211 \markcite{mcc94}(McCaughrean et al. 1994)

Is the molecular hydrogen emission consistent with a turbulent mixing layer?
It is possible to approximately determine the H$_2$ luminosity emitted 
by the blue-shifted jet. A crude estimate of the extinction can be derived 
from 1-0 Q(2) to 1-0 S(0) line ratio (Scoville et al. 1982), 
for which we obtained (for $A_{\lambda} \propto \lambda^{-1.7}$ 
\markcite{math90}, see e.g. Mathis 1990)
$A_{2.2} \approx 2.^{\rm{m}}1$. The corrected flux for the 1$-$0 S(1) 
2.12 \mum~line (for $2.^{\rm{m}}1$ of extinction) is $4.9\times 10^{-14}$ 
erg s$^{-1}$ cm$^{-2}$ for a distance of 440pc to Orion this translates 
into a luminosity $L_{2.12} \approx 1.1\times10^{30}$ erg $^{-1}$. 
If the H$_2$ 1$-$0 S(1) line carries 1/6 of the total H$_2$ luminosity 
\markcite{brad88} (Brand et al. 1988), 
then ${L_H}_2 \approx 6.85\times 10^{30}$ 
or ${L_H}_2 \approx 1.7\times10^{-3}$ \lsol.
The S(1) 2.12 \mum~flux was extracted from a section of 7.2\arcsec, so the
luminosity per unit length is ${\sl L} \approx 0.095$ \lsol~pc$^{-1}$.
The total luminosity per unit length from a turbulent mixing layer based 
on a two-dimensional model (\markcite{can91} Cant\'o \& Raga 1991; 
\markcite{nori96} Noriega-Crespo et al. 1996, eqn(1)) for a jet with a 
temperature $T_j = 10^4$ K, a radius $r_j = 10^{16}$ cm and a velocity 
$v_j = 200$ \kms, is given by 
${\sl L} = 0.056 {\left(\frac{\rm{L}_\odot}{\rm pc}\right)}
{\left(n_j\over 10^3 \rm{cm}^{-3}\right)}$, where $n_j$ is the total gas
jet density. So the jet needs a density $n_j \sim 1.7\times 10^3$ \cc~to be 
consistent with the value obtained from the spectroscopic measurements. 
The IRS2 jet H$_2$ emission, like that of HH 110, can be accounted 
for by a turbulent mixing layer.

The large radial velocities of the warm H$_2$ gas in the IRS2 jet are 
unusual, and they remind us of the apparent problem of the large tangential 
motions and shock diagnostics associated with the optical HH objects and 
jets. For several years it was difficult to reconcile the large tangential 
velocities observed in HH 1F, 
$\sim 380$~\kms~(\markcite{her81}Herbig \& Jones 1981; 
\markcite{rag90a}Raga, Mateo \& Barnes 1990; \markcite{eis94b}Eisl\"offel, 
Mundt \& B\"ohm 1994) and the exitation spectra which were indicating shock 
velocities of $\sim 100$ \kms~(see e.~g.\markcite{rag96}Raga, B\"ohm \& 
Cant\'o 1996). The HH 46/47 systems is a another example 
(\markcite{dop92}Dopita, Schwartz \& Evans 1982).
This discrepancy between tangential motions and the excitation of the optical
spectra has been interpreted in terms of a jet 
from a variable velocity source, in which relatively weak (but fast moving) 
working surfaces can be generated in the interaction region between material
ejected at different velocities (\markcite{raga90}Raga et al. 1990b).

The case for H$_2$ is probably similar, i.~e. if the radial velocities are 
indeed $\geq 100$ \kms, then direct shocks are not accelerating the warm 
molecular gas. One can imagine, however, that the gas is already moving at 
hundreds of kilometerper second (we don't know yet how this takes place 
either in optical or molecular jets), and internal weaker shocks or 
entrainment generates the H$_2$ emission.
The appealing characteristic of entrainment is that the H$_2$ excited layer 
takes a similar shape as that of the optical (ionic/atomic gas) jet.
Numerical simulations of H$_2$ jets agree with this idea, and show that is 
possible to create jets where the warm molecular gas is moving at large 
velocities without being dissociated (\markcite{sutt97}Suttner et al. 1997).

Recently, and motivated by the large radial velocities in IRS2,
we measured the proper motion of three condensations of HH 1 in the 
H$_2$ 2.121~\mum~line (Noriega-Crespo et al. 1997). We found that HH 1F 
in H$_2$ has a tangential velocity comparable to that of the ionic/atomic 
gas, i.~e. $\sim 380$~\kms. And finally, in a slightly different context, 
velocities of $40-100$~\kms~in the warm H$_2$ have been measured in OMC-1, 
where the 2.121~\mum~line profiles are unusually broad 
(see e.~g.~\markcite{nad79}Nadeau 
\& Geballe 1979; \markcite{nad82}Nadeau, Geballe \& Neugebauer 1982; 
\markcite{bra89}Brand et al. 1989).

\section{Conclusions}

Our near-infrared imaging around IRAS 05487+0255 revealed two new
highly collimated outflows visible in molecular hydrogen emission
but not seen at optical wavelengths. A single-sided jet is found to emanate 
from the bright K-band stellar source IRS1. In contrast, IRS2 K-band source
appears as a diffuse nebula dominated by scattered continuum light
from a highly obscured YSO. The YSO drives a narrow bipolar jet seen in 
molecular hydrogen emission extending more than 0.1pc from its source.
The wavelengths of the H$_2$ lines indicate relative jet velocities 
of $\sim 400$ \kms\ and imply that the jet axis is at an angle
of $\leq 45$\arcdeg~to the line-of-sight. The highly collimated 
morphology of the H$_2$ jet, the relatively large radial velocities, 
the fluxes and excitation seem to be consistent with the properties 
that are expected from a turbulent mixing layer driven by a supersonic 
atomic jet \markcite{can91}(Cant\'o \& Raga 1991; 
\markcite{tay95}Taylor \& Raga 1995).

Astrometry shows that the IRS1 and IRS2 jets are not related to 
the bright far-infrared source IRAS 05487+0255. There is, however,
a positional coincidence between the IRAS source and a faint
K-band object situated between the HH110 jet and IRS1/2. But, higher
resolution imaging in the far-IR is needed to demonstrate a physical
connection.

\acknowledgements

We thank the anonymous referee for a careful reading of the manuscript
and very useful comments.
We thank Karen Gloria and the APO staff for their support during our 
observing run. The IR spectra were obtained with the help of Marcia Rieke
and George Rieke using the FSPEC at the MMT. A. N.-C. research is supported 
by NASA Long Term Astrophysics program through a contract with the 
Jet Propulsion Laboratory (JPL) of the California Institute of Technology.

\clearpage

\begin{center}
Figure Captions
\end{center}

\figcaption[irs1_fig1]
{The region around the HH 110 jet and the IRAS 05487+0255 source
in the narrow-band filter centered at the 2.121 \mum\ H$_2$ (1$-$0 S(1)) 
line. The field is $\sim 2$\arcmin\ with north up, east to the left. The position 
of the brightest emission sources are marked (see Table 1), including 
some of the HH 110 jet knots (Davis et al. 1994b).\label{fig1}}

\figcaption[irs1_fig2]
{The region around the HH 110 jet and the IRAS 05487+0255 source
as in Figure~1, but with the continuum subtracted. The image shows
the location of the molecular emission. The bright stellar sources were
saturated in the original frames so are not properly removed in the
subtraction.\label{fig2}}

\figcaption[irs1_fig3]
{K-band spectra of both the IRS2 jet (top) and counter-jet 
(bottom). The marks correspond to the expected positions of the
strongest H$_2$ lines commonly seen in the K-band:
1-0 S(2) 2.0332~\mum, 2-1 S(3) 2.0726~\mum, 1-0 S(1) 2.1213~\mum, 
1-0 S(0) 2.2227~\mum, 2-1 S(1) 2.2471~\mum,
2-1 S(0) 2.3550~\mum, 3-2 S(1) 2.3858~\mum, 1-0 Q(1) 2.4059~\mum~and  
1-0 Q(2) 2.4128~\mum. The lower mark corresponds to 2.165 Br $\gamma$ line.
\label{fig3}}

\figcaption[irs1_fig4]
{Least square fit to the vibrational excitation temperature for the 
red-shifted (top) and blue-shifted (bottom) jets. For the red-shifted jet 
$T_{ex} = 1104\pm 67$, and for the blue-shifted jet $T_{ex} = 920\pm 156$.
\label{fig4}}

\clearpage

\makeatletter
\def\jnl@aj{AJ}
\ifx\revtex@jnl\jnl@aj\let\tablebreak=\nl\fi
\makeatother
\begin{deluxetable}{lccrr}
\tablewidth{12cm}
\tablecaption{IRAS 05487+0255 Field Positions.\label{tbl-1}}
\tablehead{ \colhead{ GNRB\tablenotemark{1} } &
\colhead{Identification} & \colhead{Other} &
\colhead{$\alpha$ (1950)} & \colhead{$\delta$ (1950)}\\
\colhead{} &  \colhead{} & \colhead{Names} & \colhead{($^h$ $^{m}$ $^{s}$)} 
& \colhead{(\arcdeg~\arcmin~\arcsec)}
}
\startdata
 1 & Star &  & 5 48 49.05  &  2 56 08.3 \nl
 2 & HH110 knot & A1\tablenotemark{2}   & 5 48 48.47  &  2 54 56.4 \nl
 3 & HH110 knot & A2\tablenotemark{2}   & 5 48 48.40  &  2 54 52.8 \nl
 4 & HH110 knot & B1\tablenotemark{2}   & 5 48 48.37  &  2 54 49.1 \nl
 5 & HH110 knot & B2\tablenotemark{2}   & 5 48 48.30  &  2 54 44.4 \nl
 6 & HH110 knot & E1\tablenotemark{2}   & 5 48 47.53  &  2 54 19.6 \nl
 7 & IRAS source? & & 5 48 47.03  &  2 55 11.1 \nl
 8 &      & & 5 48 46.27  &  2 54 11.7 \nl
 9 & Counter-jet & IRS2\tablenotemark{2}  & 5 48 45.39  &  2 55 28.9 \nl
10 & Source & IRS2\tablenotemark{2} & 5 48 45.31  &  2 55 24.8 \nl
11 & Jet & IRS2\tablenotemark{2} & 5 48 45.46  &  2 55 17.5 \nl
12 & Emission knot & IRS1\tablenotemark{2} & 5 48 45.47  &  2 54 48.1 \nl
13 & Emission knot & IRS1\tablenotemark{2} & 5 48 45.61  &  2 54 43.0 \nl
14 &      &  & 5 48 45.52  &  2 54 27.5 \nl
15 & Source & IRS1\tablenotemark{2} & 5 48 45.33  &  2 55 01.1 \nl
16 & Star  &   & 5 48 45.00  &  2 54 59.6 \nl
17 & Star & Ref star 1\tablenotemark{3} & 5 48 43.46  &  2 54 43.2 \nl
\enddata
\tablenotetext{1}{We recommend the standard IAU designation convention be used,
for example the second object above: HH110:GNRB 2}
\tablenotetext{2}{Davis et~al. (1994b)}
\tablenotetext{3}{Position from Reipurth \& Olberg 1991}
\end{deluxetable}

\clearpage

\def \flux{\rm{erg} \rm{s}$^{-1}$ \rm{cm}$^{-2}$}
\def \inten{\rm{erg} \rm{s}$^{-1}$ \rm{cm}$^{-2}$ \rm{sr}$^{-1}$}
\def \cm2{\rm{cm}$^{-2}$}
\makeatletter
\def\jnl@aj{AJ}
\ifx\revtex@jnl\jnl@aj\let\tablebreak=\nl\fi
\makeatother
 
\begin{deluxetable}{llrrcc}
\tablecaption{IRS2 H$_2$ Bipolar Jet Spectra.\label{tbl-2}}
\tablehead{
\colhead{$\lambda_{rest}$(\mum)} & \colhead{$\lambda_{obs}$(\mum)} 
& \colhead{Flux\tablenotemark{2}}
& \colhead{Intensity\tablenotemark{3}} 
& \colhead{E(cm$^{-1}$)\tablenotemark{4}}
& \colhead{ln(N(v$^\prime$, J$^\prime$)/g$_{tot}$)}
}
\startdata
\hfil North (Counter-jet)\hfil  &       &    &     &     &     \nl
1-0 S(2) 2.0332 & 2.034 &  7(3) &  8(3) & 5271.36 & 30.9(0.3) \nl
1-0 S(1) 2.1213 & 2.1226\tablenotemark{1}& 26(2) & 28(2) 
& 4831.41 & 31.5(0.1) \nl
1-0 S(0) 2.2227 & 2.224 & 23(5) & 25(5) & 4497.82  & 33.3(0.2) \nl
2-1 S(1) 2.2471 & 2.249 &  2(2) &  2(2) & 8722.70  & 27.6(0.7) \nl
1-0 Q(1) 2.4059 & 2.405 & 30(5) & 32(5) & 4273.75  & 32.4(0.1) \nl
1-0 Q(2) 2.4128 & 2.410 &  5(2) &  5(2) & 4497.82  & 31.5(0.3) \nl
\hfil South (Jet)\hfil  &       &       &         &    &    \nl
1-0 S(2) 2.0332 & 2.032 & 42(10)& 25(5) & 5271.36 & 32.1(0.2) \nl
1-0 S(1) 2.1213 & 2.1194\tablenotemark{1} & 72(5) & 42(3) & 4831.41 
& 32.0(0.1) \nl
1-0 S(0) 2.2227 & 2.220 & 21(6) & 12(3) & 4497.82 & 32.5(0.2) \nl
2-1 S(1) 2.2471 & 2.246 &  9(4) &  5(2) & 8722.70 & 28.7(0.3)\nl
1-0 Q(1) 2.4059 & 2.404 & 92(10)& 54(5) & 4273.75 & 33.0(0.1) \nl 
1-0 Q(2) 2.4128 & 2.410 & 24(8) & 14(4) & 4497.82 & 32.5(0.3) \nl
\tablenotetext{1}{with a uncertainty of $\pm 0.0005$ \mum}
\tablenotetext{2}{Units of $10^{-16}$ \flux}
\tablenotetext{3}{Units of $10^{-6}$ \inten}
\tablenotetext{4}{From Dabrowski, p1653, Table 5 (1994)}
\enddata
\end{deluxetable}
 
\clearpage






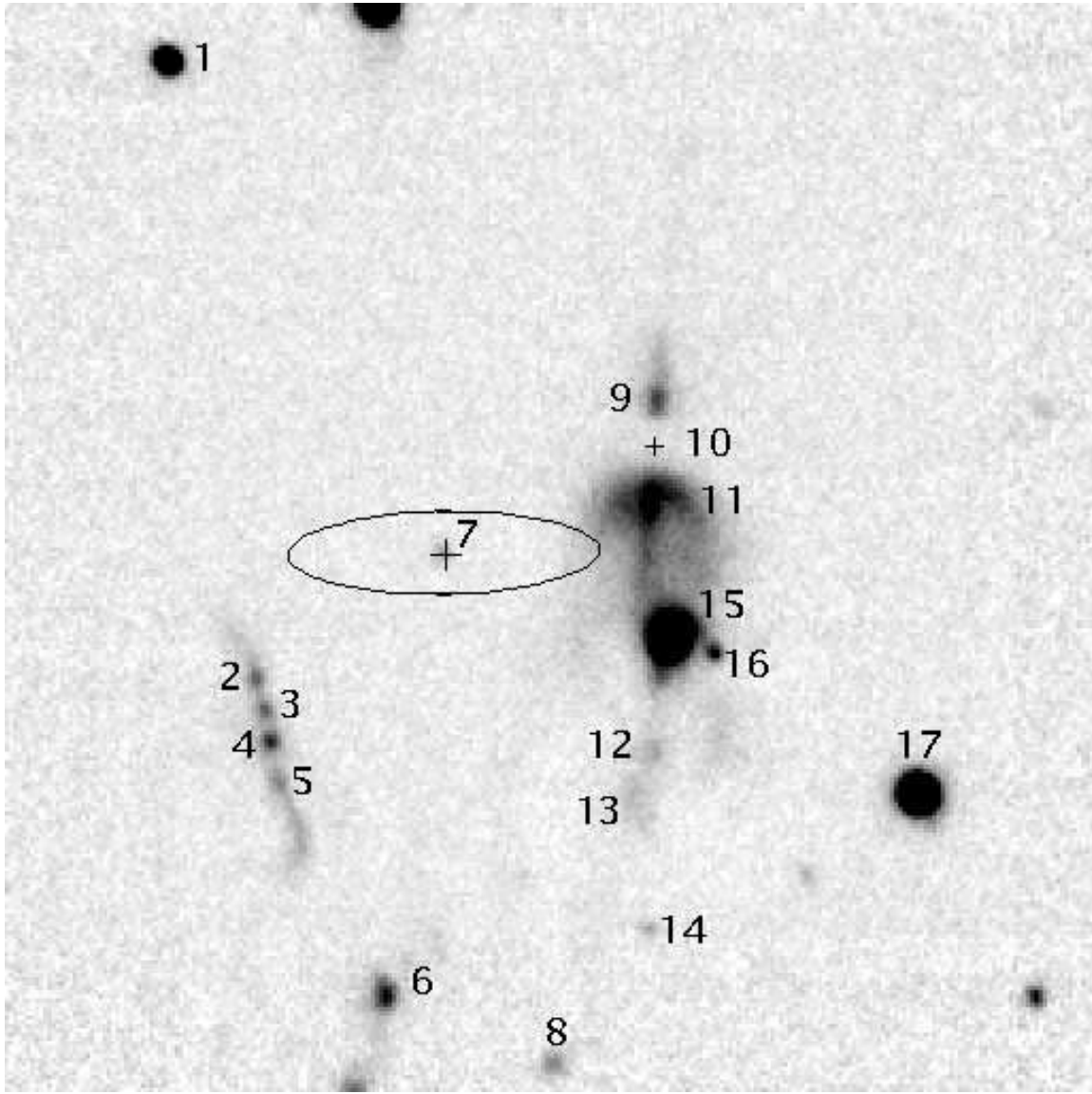
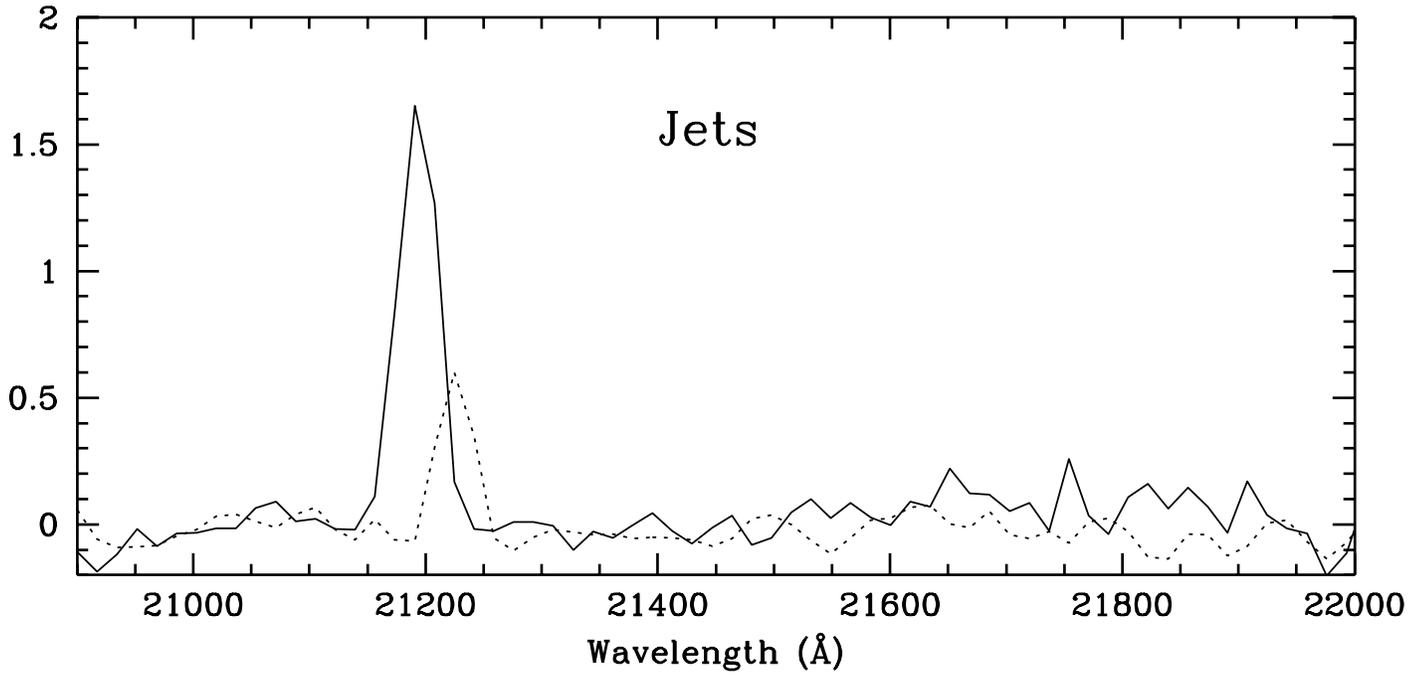
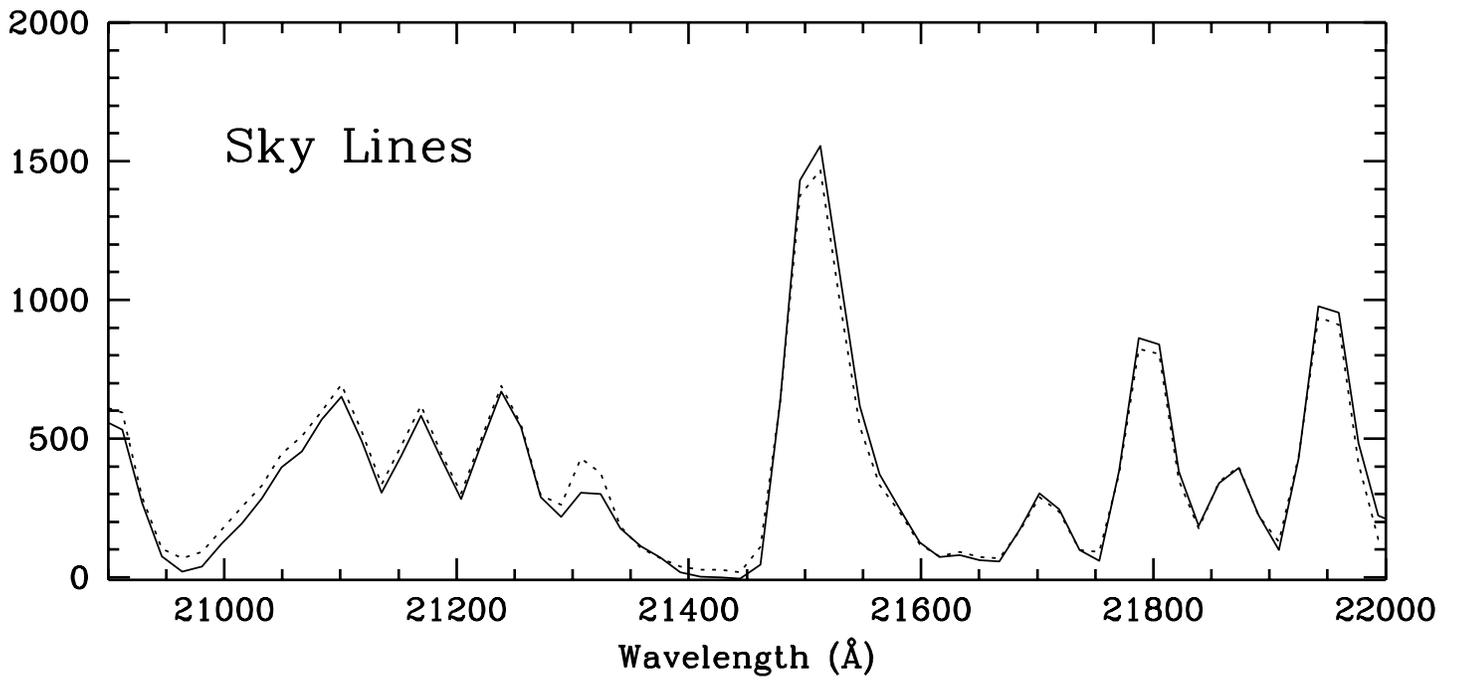
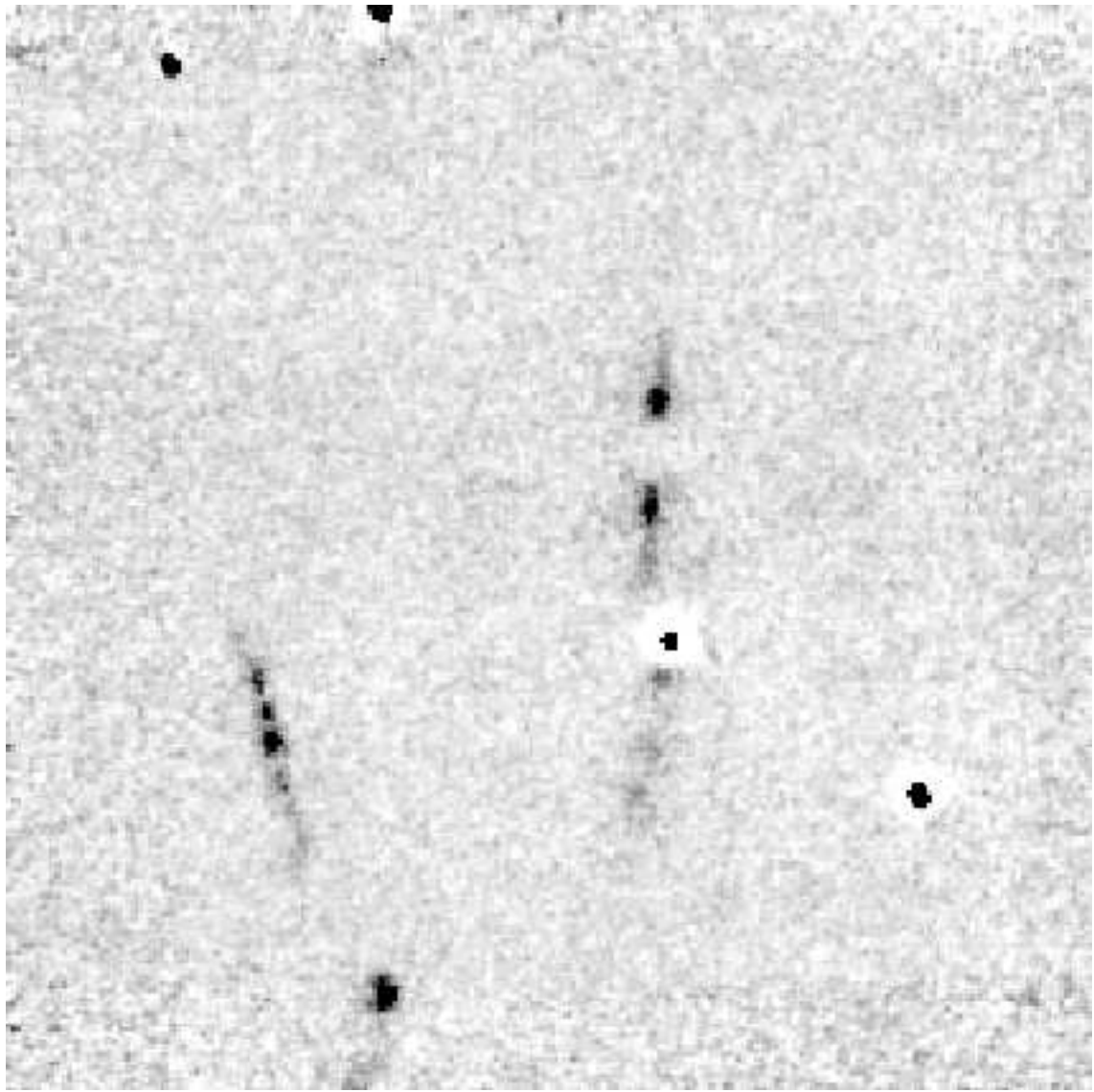
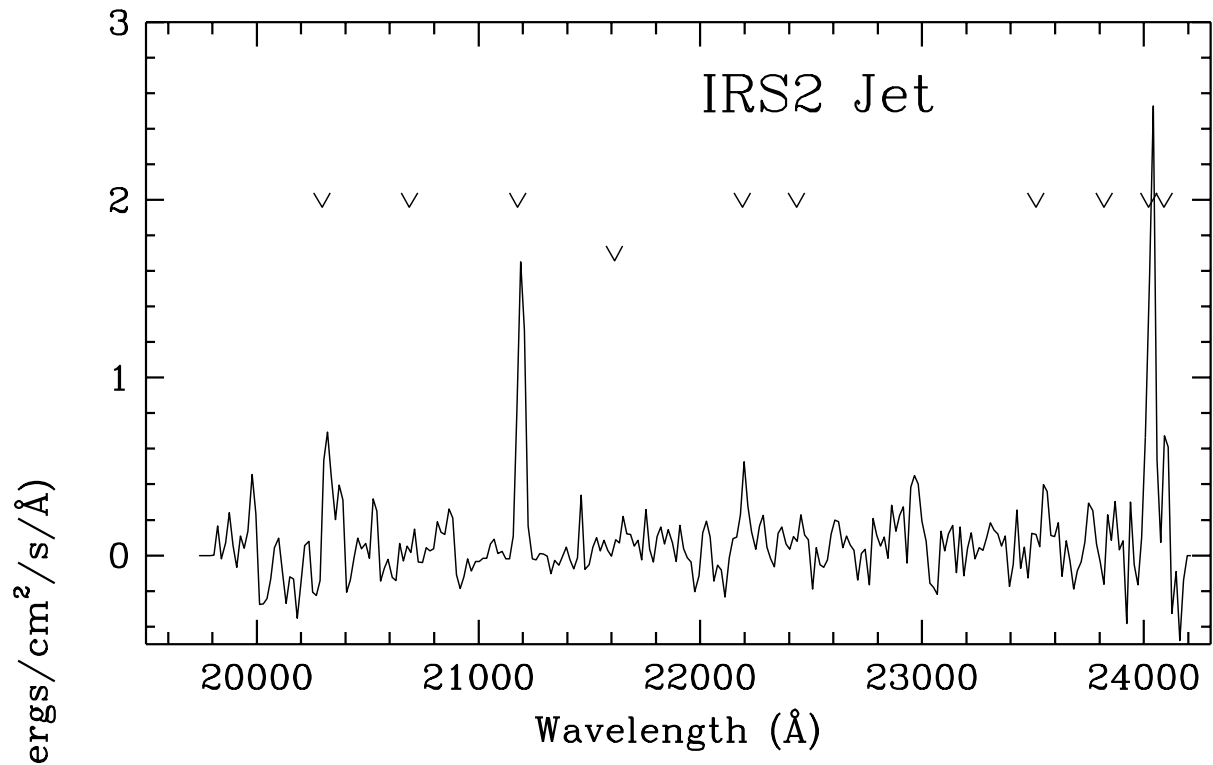
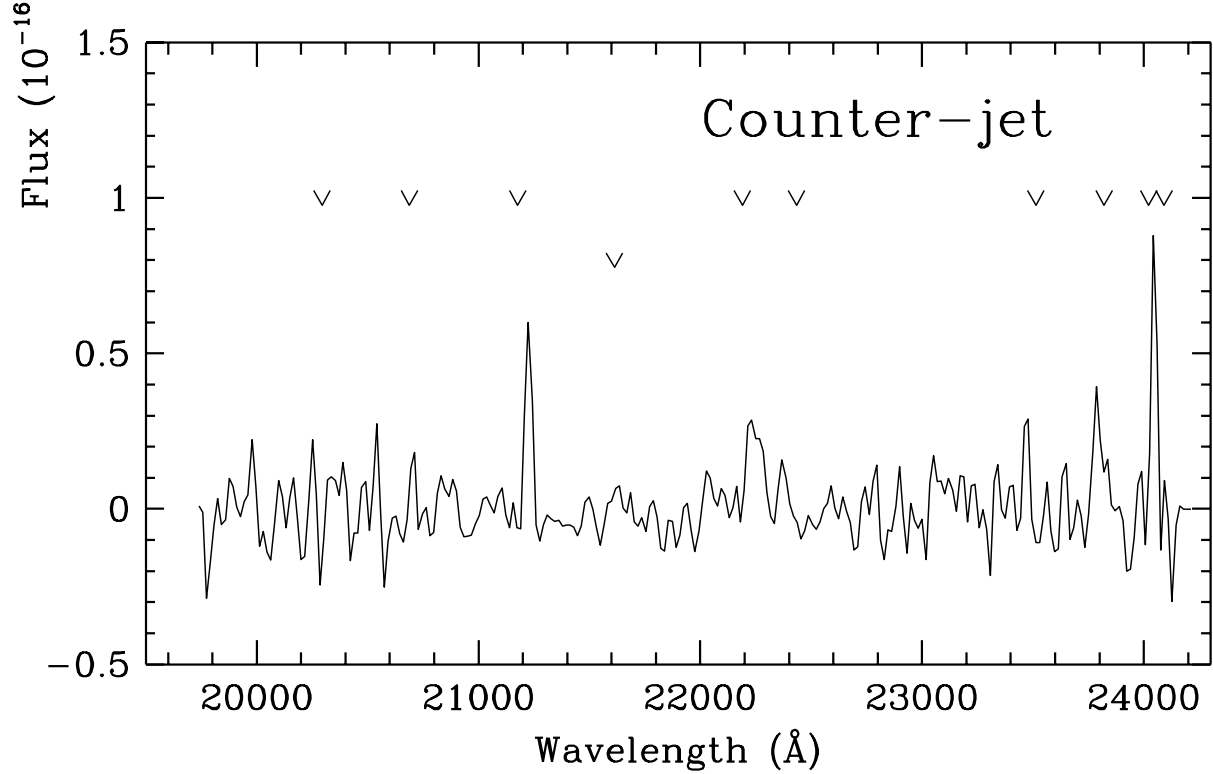
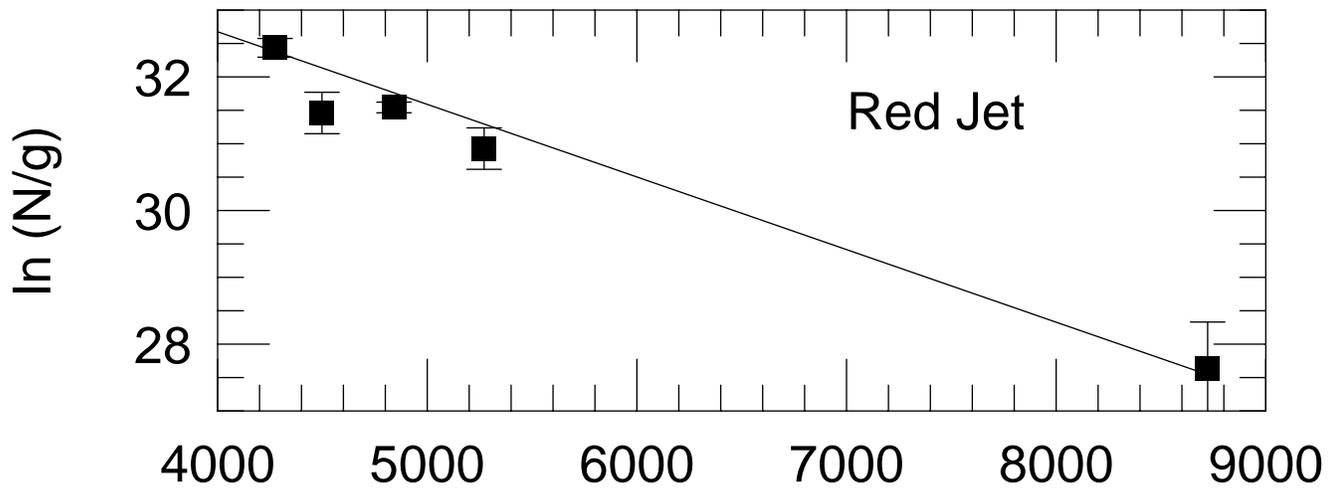
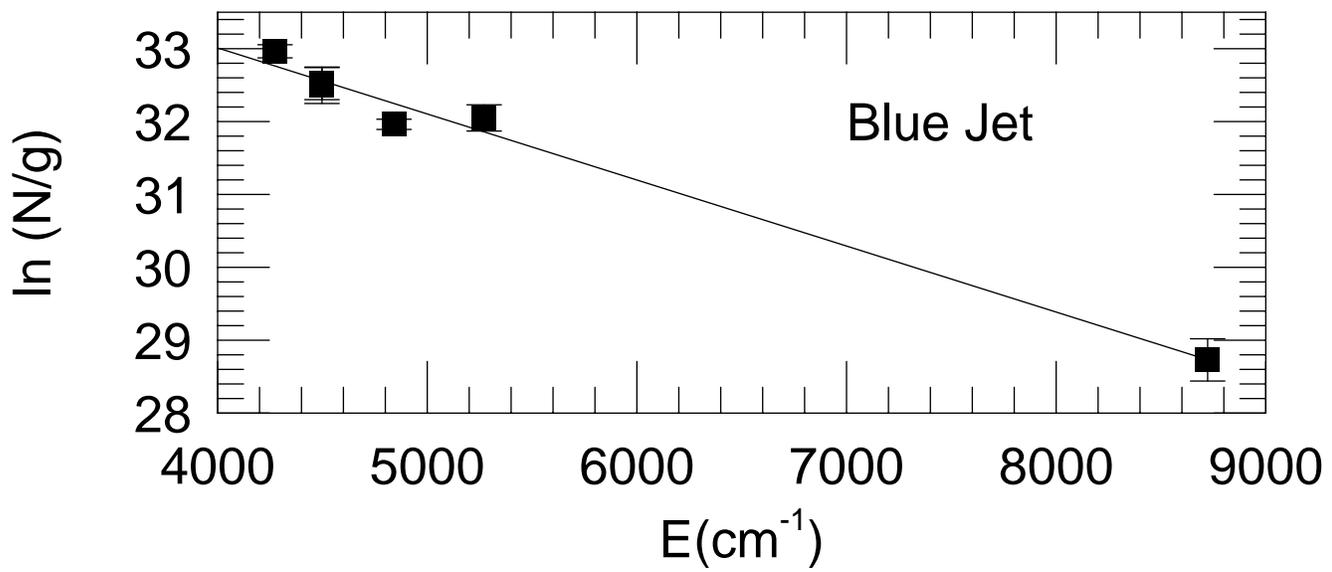
\end{document}